\documentclass[11pt,cite,epsfig,psfrag]{article}
\usepackage[utf8]{inputenc}
\usepackage{placeins}
\usepackage{geometry}
 \usepackage{graphicx}
 \usepackage{subfig}
  \usepackage{amsmath} 
  \graphicspath{{Images_matrix/}}
  \usepackage[usenames, dvipsnames]{color}
\usepackage{wrapfig}
\usepackage{boxedminipage}
\usepackage{multirow}

\usepackage{fullpage}
\usepackage{amsmath}
\usepackage{amssymb}
\usepackage{graphicx}
\usepackage{mathrsfs}
\usepackage{wrapfig}
\usepackage{boxedminipage}
\usepackage{epsfig}
\usepackage{setspace}
\usepackage{float}
\usepackage{hyperref}
\usepackage{enumerate}
\usepackage{cite}
\usepackage[font=footnotesize,labelfont=rm]{caption}

\usepackage[numbers,sort&compress]{natbib}

\def\be{\begin{equation}}
\def\ee{\end{equation}}
\def\bea{\begin{eqnarray}}
\def\eea{\end{eqnarray}}

\numberwithin{equation}{section}

 \newcommand{\RN}[1]{%
   \textup{\uppercase\expandafter{\romannumeral#1}}%
 }
\begin{document}

\thispagestyle{empty}
%\baselineskip 20pt
%\rightline{IP/BBSR/2005-12}
%\rightline{\tt hep-th/yymmnnn}

\vskip 2cm

\begin{center}
{\Large \bf Cardy-Verlinde formula from boundary matrix model}
\end{center}

\vskip .2cm

\vskip 1.2cm

\centerline{ \bf   Pavan Kumar Yerra \footnote{pk11@iitbbs.ac.in}, Chandrasekhar Bhamidipati\footnote{chandrasekhar@iitbbs.ac.in} and Sudipta Mukherji\footnote{mukherji@iopb.res.in}
}

\vskip 7mm 
\begin{center}{ $^{1},^{3}$ Institute of Physics, Sachivalaya Marg, \\ Bhubaneswar, Odisha, 751005, India \\ and\\Homi Bhabha National Institute, \\Training School Complex, \\Anushakti Nagar, Mumbai, 400085, India}	
\end{center}

\begin{center}{ $^{2}$ School of Basic Sciences\\ 
Indian Institute of Technology Bhubaneswar \\ Bhubaneswar, Odisha, 752050, India}
\end{center}

\vskip 1.2cm
\vskip 1.2cm
\centerline{\bf Abstract}
\vskip 0.5cm

Cardy-Verlinde (CV) formula relates the entropy of a strongly coupled 
conformal field theory (CFT) possessing AdS dual, at finite temperature, 
to its total energy (with an appropriate insertion of additional 
internal energy
for charged systems) and Casimir energy. While the CV formula has been 
checked for various CFTs, in the present work, we intend to verify it 
directly, by exploiting a phenomenological matrix model, which is 
believed to qualitatively capture features of strongly coupled ${\cal N} 
= 4, SU(N)$ Super Yang-Mills theory on $S^3$ at finite temperature and 
chemical potential at large $N$

\noindent

\vskip 0.5cm
\noindent

%\begin{quote}
%\noindent
%\end{quote}
\newpage
\setcounter{footnote}{0}
\noindent

\baselineskip 15pt
\section{Introduction}
A couple of decades ago, Verlinde observed that the entropy of a
strongly coupled conformal field theory (CFT), possessing AdS dual,
satisfies a Cardy-like formula at finite temperature~\cite{Verlinde:2000wg}. This
later came to be known  as the Cardy-Verlinde (CV) formula. For
CFTs on $R \times S^n$, it reads
\begin{equation}
	S = \frac{2\pi R}{n} {\sqrt{E_c (2 E -E_c)}}.
	\label{eq:C_V}
\end{equation}
Here, $S$ is the entropy associated with the CFT, $R$ is the radius of
the $S^n$, $E$ is the total energy and $E_c$ is the
subextensive part of it. This relation was shown to arise
holographically from the thermodynamic quantities associated with the
AdS-Schwarzschild black hole. The CV formula drew immediate
attention to many as it satisfies a Bekenstein like entropy bound in $n$
dimensions, namely,
\begin{equation} S \le \frac{2\pi R E}{n}.
	\label{eq:B_Bound}
\end{equation}
Subsequently, the formula found its generalization to many other CFTs.
One example that will be important for our purpose is the one dual to an
electrically charged AdS black hole. The entropy was then found to obey~\cite{cai2001cardyverlinde}
\begin{equation}
	S = \frac{2\pi R}{n} {\sqrt{E_c [2 (E- Eq) -E_c]}},
	\label{eq:C_V_charged}
\end{equation}
where $E_q$ is the electrostatic potential energy.\\

\noindent
While a direct computation~\cite{Kutasov:2000td} of the entropy of a weakly
coupled CFT is shown to violate the bound (\ref{eq:B_Bound}),
for a strongly coupled CFT, a similar analysis is missing. In this work
we wish to take a
step along this direction. Generally, to describe a gauge theory on $S^3
\times S^1$ at zero coupling, one writes down an effective action
involving
the  vev of the Wilson-Polyakov loop variable on $S^1$ as the only light
degree of freedom~\cite{Sundborg:1999ue,Aharony:2003sx,Alvarez-Gaume:2005dvb,Basu:2005pj,Dey:2006ds,Dey:2007vt,Dey:2008bw,Chandrasekhar:2012vh}. Since it is hard
to compute such an action at finite Yang-Mills coupling, one may resort
to phenomenological model belonging to the same universality class of
the boundary theory. Indeed, such a phenomenological model
was proposed in~\cite{Alvarez-Gaume:2005dvb}. It contains two parameters
which  depend on the temperature and $\lambda$, the 't Hooft coupling.
In~\cite{Alvarez-Gaume:2005dvb,Yerra:2023hui}, these dependences were found
by
exploiting the data arising from the dual supergravity description.
Subsequently, the model was
generalized  to incorporate the $R$-charge emerging from the charged AdS
black hole~\cite{Basu:2005pj,Yerra:2023hui}. This is done by allowing
the model parameters to depend on the chemical potential as well. Armed
with the results of these investigations,
in the present work, we check if the model is consistent with the CV
formula. To this end, we compute the left and the right hand sides of
the
equation (\ref{eq:C_V}) directly from the model and find qualitative
agreement. A similar agreement follows for the
charged case as well when compared against the equation
(\ref{eq:C_V_charged}). Due to the complexity of the model, at places
analytical handle becomes difficult and we had to resort to the
numerical computations.\\

\noindent
The organization of this paper is as follows. In section-(\ref{one}), we start
with the AdS-Schwarzschild black hole. In subsection-(\ref{section:boundary for sch} ), we briefly
review
the corresponding matrix model living on the boundary. Subsection-(\ref{2.2})
contains a direct check on the CV formula arising from 
the model. Subsequently, in section-(\ref{3}), working in the canonical ensemble, we generalize our computation
of the previous section to the electrically charged black hole case. In subsection-(\ref{canonicalboundary}), we  review the extension
of the matrix model to include non-zero chemical potential.
The computation leading to the
agreement with the CV formula is presented in the next subsection.
Section-(\ref{conclusion}) contains some remarks. We end the paper with two
appendices where the thermodynamic properties associated with the bulk
black holes are summarized.

%%%%%%%%%%
\section{Cardy-Verlinde formula from boundary matrix model: zero chemical potential} \label{one}
%%%%%%%%%%%%
In this section, we start by considering the case of a conformal field theory (CFT) represented by a matrix model dual to the Schwarzschild-AdS$_5$ black hole on the gravity side. We 
compute the  entropy of the CFT from the effective potential of the matrix model. Considering the CFT in a finite volume, the Casimir energy can be obtained. Then, we numerically verify the Cardy-Verlinde formula  relating the entropy  with its energy and Casimir energies.

%%%%%%%%%%
\subsection{Boundary matrix model: zero chemical potential}\label{section:boundary for sch} 
\begin{figure}[h!]
	
	% \begin{wrapfigure}{r}{0.43\textwidth}
	%\begin{center}
	{\centering
		
		\subfloat{\includegraphics[width=2.8in]{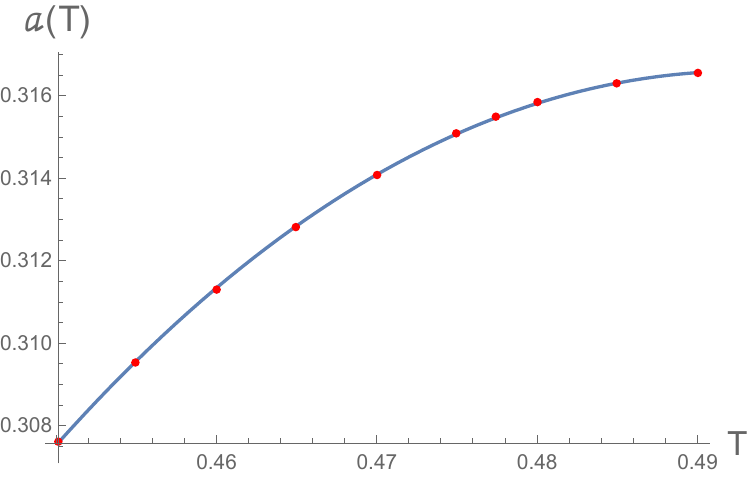}}\hspace{1cm}	
		\subfloat{\includegraphics[width=2.8in]{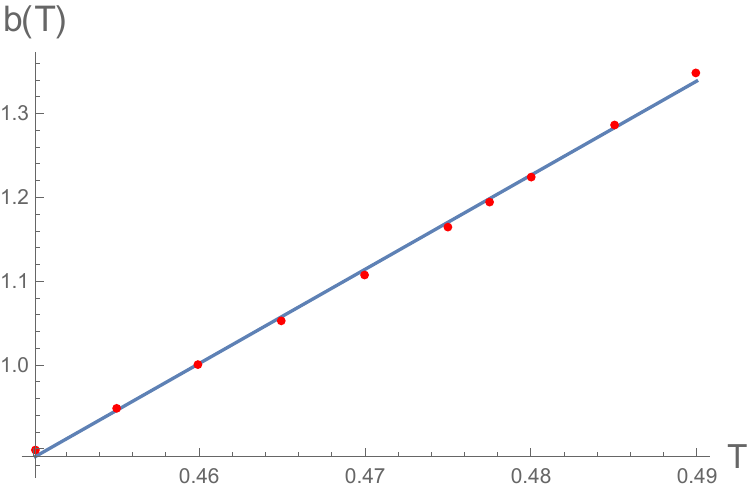}}	
		
		\caption{\footnotesize Plots show the temperature dependence of the parameters $a(T)$ and $b(T)$ for the phenomenological $(a,b)$ matrix model at zero chemical potential. The red dots denote the data points, which are fitted by the blue colored curves. The fitting formulas are $a(T) =c_1 \text{log}(T) +c_2 T +c_3$, and $b(T) =c_4 T +c_5$, $(\text{where,}\, c_1=2.257, c_2=-4.579, c_3=4.171, c_4=11.239, c_5=-4.169.)$.}
		\label{fig:sch_boundary_a_b_plots}	}
	
\end{figure}

\begin{figure}[h!]

	% \begin{wrapfigure}{r}{0.43\textwidth}
	%\begin{center}
	{\centering
		
		\subfloat[]{\includegraphics[width=2.8in]{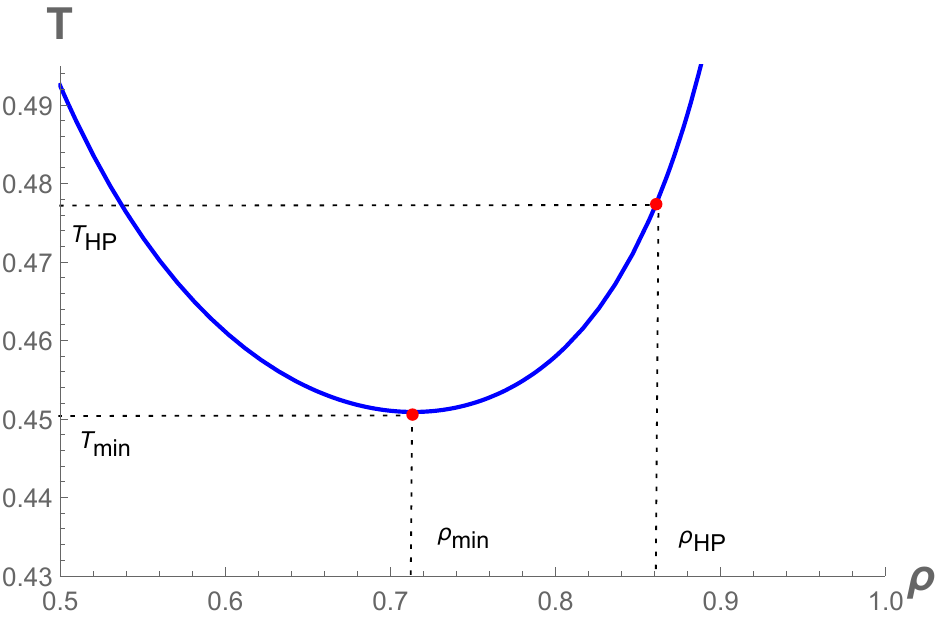}	\label{fig:cv_sch_matrix_T}}\hspace{0.5cm}
		\subfloat[]{\includegraphics[width=3in]{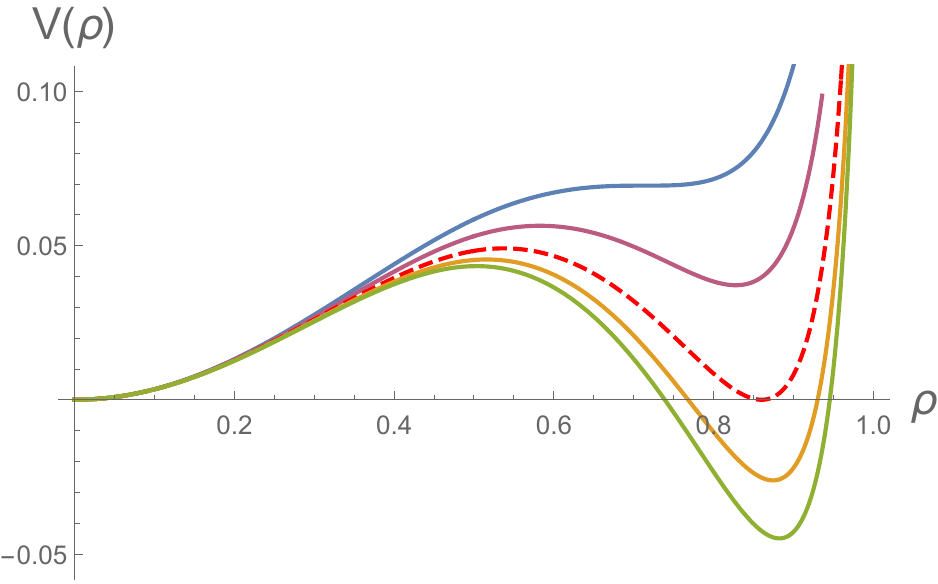}	\label{fig:sch_boundary_effective_potential_plot}}

		\caption{\footnotesize Case of $(a,b)$ matrix model at zero chemical potential: (a) Nature of $T(\rho)$ qualitatively resembles the behaviour of the equation of state of the bulk (cf. Fig.~\ref{fig:cv_sch_bh_T} in Appendix-A). (b) The effective potential $V(\rho)$ at various temperatures $T$ is shown. Temperature of the curves increases from top to the bottom. Blue curve is for $T_{\rm min}$, where as the dashed red curve showing the HP transition is for $T_{\rm HP}$, with $a=0.3155$ and $b=1.194$, $(\rho_{\rm min},\rho_{\rm HP}) = (0.714, 0.861)$.}
	}	
\end{figure}

%%%%%%%%%%%%%
Our starting point is the phenomenological matrix theory, called the $(a,b)$ model, which has been shown to capture, among other aspects, the deconfinement phase transition of $\mathcal{N}=4$, $SU(N)$ guage theory 
in the large $N$ limit. We only describe this model briefly, as details are available in~\cite{Alvarez-Gaume:2005dvb,Aharony:2003sx}. We also closely follow the set up in~\cite{Dey:2006ds,Dey:2008bw,Chandrasekhar:2012vh,Yerra:2023hui}, well suited for our purposes. 
\noindent
To this end, the partition function of the SYM theory is written as a matrix integral over the 
effective action  consisting of the operator corresponding to the Wilson-Polyakov loop, i.e., $(\text{tr}U)/N$ as
\begin{equation}
	Z(\lambda, T)=  \int dU e^{S_\text{eff}(U)}.
\end{equation}
Here, $U = P\, \text{exp}(i \int_{0}^{\beta} A d\tau)$ is the $U(N)$ unitary matrix, where $A(\tau)$ stands for the zero mode (of the time component) of the
gauge field on $S^3$. Generally, the action $S_\text{eff}(U)$ turns out to be a polynomial in the traces of $U$ and its powers, which 
are allowed by imposing the $Z_N$ symmetry. Now, by truncating the above action and retaining only a couple of terms, one ends up with the aforementioned phenomenological $(a,b)$ model, which takes the form,
\begin{equation}\label{eq: a_b_matrixmodel_for_sch}
	Z(a, b) = \int dU \, \text{exp}[ a (\text{tr}U \, \text{tr}U^{\dag}) + \frac{b}{N^2} (\text{tr}U \, \text{tr}U^\dag)^2].
\end{equation}
Here, $a$ and $b$ are the key parameters of the model which contain nontrivial dependence on the temperature $T$ and 
the ’t Hooft coupling $\lambda$. Furthermore, the effecting potential ensuing from this set up can be expressed in terms of the order parameter $\rho$ ( which is the expectation value of the Polyakov loop $ \frac{1}{N} \langle\text{tr} \, U\rangle$) that characterizes the deconfined phase of the gauge theory as:
\begin{eqnarray}\label{eq:Veff_sch_boundary}
	V(\rho) &=& \frac{1-a}{2}\rho^2 -\frac{b}{2} \rho^4      \hspace{4cm} \text{for} \quad 0 \leq \rho \leq \frac{1}{2}    \\
	&=& -\frac{a}{2} \rho^2 -\frac{b}{2} \rho^4 -\frac{1}{4} \text{log}[2(1-\rho)] +\frac{1}{8}  \hspace{0.8cm}  \text{for} \quad \frac{1}{2} \leq \rho \leq 1. \label{eq:eff_potential for rho > 1/2}
\end{eqnarray}  
Here, $\rho^2 = (\frac{1}{N^2}) \text{tr}\, U \text{tr}\, U^\dag$ and the saddle point equation is
\begin{eqnarray}
	a\rho +2b\rho^3 &=& \rho   \hspace{2cm} \text{for} \quad 0 \leq \rho \leq \frac{1}{2}    \\
	&=& \frac{1}{4(1-\rho)}  \hspace{0.8cm}  \text{for} \quad \frac{1}{2} \leq \rho \leq 1. \label{eq:saddle pt eq for rho >1/2 }
\end{eqnarray} 
The parameters satisfy the bounds $a <1$ and $b>0$.  The dependence of the parameters $a(T)$ and $b(T)$ on the temperature $T$ can be obtained numerically making use of the bulk data, as in~\cite{Alvarez-Gaume:2005dvb,Dey:2006ds,Yerra:2023hui}. The result is captured in Fig.~\ref{fig:sch_boundary_a_b_plots}.
\vskip 0.2cm
\noindent
Now, utilizing the fitting curves for $a(T)$ and $b(T)$, the saddle point equation~\eqref{eq:saddle pt eq for rho >1/2 } can be solved, yielding the following expression for temperature:
 \begin{equation}\label{eq:sch_boundary_T}
 T(\rho) = \frac{c_1}{(c_2 + 2c_4 \rho^2)}\, \text{ProductLog} \Big[ \frac{e^{\frac{1-4c_3 (1-\rho)\rho - 8c_5 (1-\rho) \rho^3}{4c_1 \rho (1-\rho)}}  (c_2 + 2c_4\rho^2)}{c_1} \Big].
 \end{equation}
\noindent As seen from Fig.~\ref{fig:cv_sch_matrix_T}, $T(\rho)$ mimics the behaviour of the equation of state of the theory in the bulk (see Fig.~\ref{fig:cv_sch_bh_T} in Appendix-A).  
\vskip 0.2cm
\noindent
Now, let us discuss the nature of the effective potential $V(\rho)$, which has been plotted in Fig.~\ref{fig:sch_boundary_effective_potential_plot} for various temperatures. In the $\rho > 1/2$ region, the phase structure of the bulk system emerges from the saddle points of $V(\rho)$~\cite{Alvarez-Gaume:2005dvb}.  At any temperature above a certain minimum temperature $T_{\rm min}$, one gets two saddle points corresponding to large black holes (LBH) and small black holes (SBH), with the former being stable and the later unstable. $T=T_{\rm min}$ is the nucleation point for the creation of this black hole pair. Minimum occurring at $\rho=0.861$ denotes the HP transition temperature in the bulk, which corresponds to the deconfining temperature of the boundary theory. The minima occuring at  $\rho = 0$ stands for the thermal $\text{AdS}_5$ phase. It is useful to note that the supergravity description is not valid in the $\rho \leq 1/2 $ region and at $\rho=1/2$ the matrix model exhibits a third order Grass-Witten transition~\cite{Gross:1980he}, which has a natural intepretation as a Horowitz-Polchinski point in string theory~\cite{Horowitz:1996nw}.
\vskip 0.2cm
\noindent
Since, here the effective potential $V(\rho)$ plays the role of off-shell free energy of the bulk theory (see equation-(\ref{fe1}) in Appendix-A), in what follows, we use it to derive other thermodynamic quantities, as is generally done in standard treatment.
The on-shell free energy $F$ can now be obtained, using the fitting curves for $a(T)$, $b(T)$, and on substituting $T(\rho)$ in  the effective potential $V(\rho)$ in equation~\eqref{eq:eff_potential for rho > 1/2}, as
\begin{equation}
F = -\frac{\rho^2}{2}(c_1 \text{log}T +c_2 T+ c_3) -\frac{\rho^4}{2}(c_4 T+ c_5)-\frac{1}{4}\text{log}[2(1-\rho)]+\frac{1}{8}.\label{eq:on-shell_F}
\end{equation} 
\noindent
The computation of the entropy $S$ of the CFT (assuming a relation of the form $dF=-SdT$ analogous to bulk free energy) turns out to be
\begin{equation}\label{eq:sch_boundary_S}
S=-\frac{\partial F}{\partial T} = -\frac{\partial F /\partial \rho}{\partial T/ \partial \rho}.
\end{equation}  
Although, the above can be evaluated straightforwardly, the expression is large and hence we do not present it here.
%%%%%%%%%%
\subsection{Cardy-Verlinde formula} \label{2.2}
\begin{figure}[h!]
	
	% \begin{wrapfigure}{r}{0.43\textwidth}
	%\begin{center}
	{\centering
		
		{\includegraphics[width=2.8in]{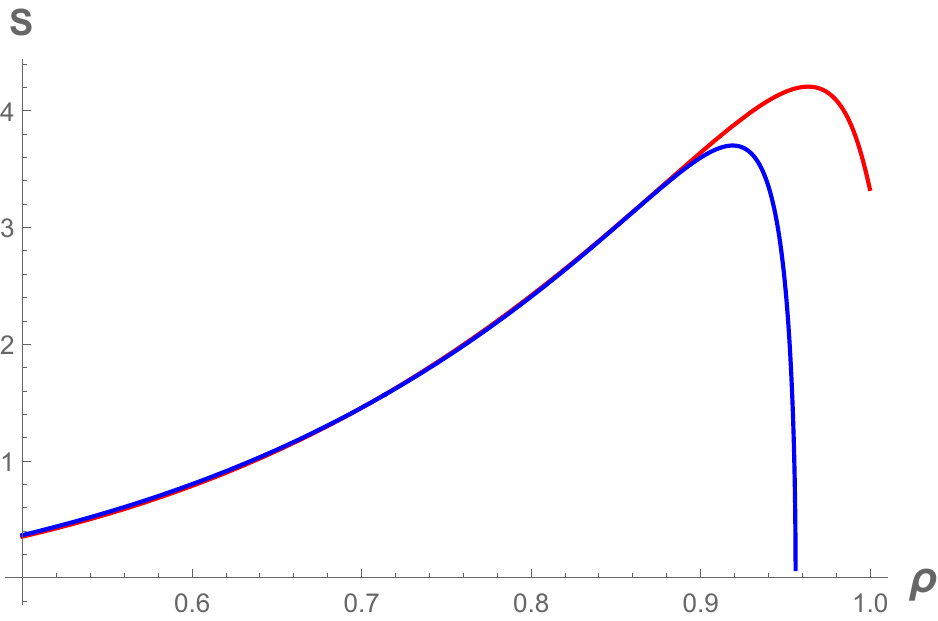}}	
		\caption{\footnotesize For $(a,b)$ matrix model with zero chemical potential: Red curve is for the LHS  and the blue curve is for the RHS of the equation~\eqref{eq:cv_sch}.}
		\label{fig:cv_sch_matrix_f}	}
	
\end{figure}

%%%%%%%%%%%%%
Now, to derive the Cardy-Verlinde (CV) formula, we closely follow the set up in~\cite{Cappiello:2001tf}. We write  the boundary metric, from the large $r$ behaviour of the bulk metric~\eqref{eq:bulk_sch_metric}, as
\begin{equation}
ds^2=\frac{r^2}{l^2} (-dt^2+l^2d\Omega_3^2).
\end{equation}
After scaling the time parameter to $\tau =\frac{R}{l}t$, the above metric can be recast as
\begin{equation}\label{eq:sch_scaled_metric}
ds^2=\frac{r^2}{R^2} (-d\tau^2+R^2d\Omega_3^2),
\end{equation}
which is conformally equivalent to $\mathbb{R} \times S^3$, where $S^3$ now has radius $R$.
According to the holographic principle, the temperature $\bar{T}$ and the  free energy $\bar{F}$ of the CFT can be obtained by scaling the respective expressions in~\eqref{eq:sch_boundary_T} and~\eqref{eq:on-shell_F} as~\cite{,Cappiello:2001tf}:
\begin{eqnarray}
\bar{T} &=& \frac{1}{\bar{\beta}}=\frac{l}{R}T, \\
\bar{F} &=& \frac{l}{R}F,
\end{eqnarray}
\noindent with the CFT volume taken to be $\bar{V}= \omega_3 R^3.$   Assuming the free energy to be $\bar{F} = \bar{E}- \bar{T} \bar{S}$, we derive the energy $\bar{E}$ as a function of temperature $\bar{T}$ and volume $\bar{V}$ as  
\begin{equation}
	\bar{E} = \frac{\partial(\bar{\beta} \bar{F})}{\partial\bar{\beta}} \Big\vert_{\rm \bar{V}} = \frac{\partial(\bar{\beta} \bar{F})/\partial \rho}{\partial\bar{\beta}/\partial{\rho}} \Big\vert_{\rm R},
\end{equation}
while, the entropy $\bar{S}$ is
\begin{equation}
\bar{S}= \bar{\beta}(\bar{E}-\bar{F})=S.
\end{equation}
This shows that the entropy of the CFT does not scale with $R$. Now, the pressure $\bar{P}$ can be defined as
\begin{equation}
\bar{P}=-\frac{\partial \bar{F}}{\partial \bar{V}} \Big \vert_{\rm \bar{\beta}} = -\frac{\partial \bar{F}(\bar{\beta}, \rho)/ \partial \rho}{\partial \bar{V}/ \partial \rho} \Big \vert_{\rm \bar{\beta}}, 
\end{equation}
\noindent which satisfies the equation of state $\bar{E}=3\bar{P}\bar{V}.$  The Gibbs free energy  is $\bar{G}= \bar{F}+ \bar{P}\bar{V}$, and thus the Casimir energy can be defined by $\bar{E}_c = 3\bar{G}$~\cite{Verlinde:2000wg}. The expressions for the aforementioned thermodynamic quantities can be obtained, but are quite lengthy. Hence, we directly show our result in Fig.~\ref{fig:cv_sch_matrix_f}, which confirms validity of the following Cardy-Verlinde formula given as~\cite{Verlinde:2000wg}:
\begin{equation}\label{eq:cv_sch}
\bar{S}=\frac{2\pi R}{3}\sqrt{\bar{E}_c (2\bar{E}-\bar{E}_c )}.
\end{equation} 
There is however a  mismatch of the curves  around $\rho =1$ in Fig.~\ref{fig:cv_sch_matrix_f}, which we attribute to some statistical errors and the approximations which were considered in the current model. 
%We note here that, as the resulting expressions in the above computations are huge, we are not writing them.

%%%%%%%%%%%%
\section{Cardy-Verlinde formula from boundary matrix model: non-zero chemical potential}  \label{3}
%%%%%%%%%%%%

In this section, we  consider a boundary  matrix model with  chemical potential dual to Reissner-Nordstrom-AdS$_5$  black holes in canonical (fixed charge) ensemble. 
Using the effective potential of the matrix model, we perform a computation similar to the one in the previous section. We also numerically verify the Cardy-Verlinde formula, which now contains electrostatic potential energy coming from the charge, in addition to the Casimir energy present earlier.  
%%%%%%%%%%%%%%%%%%%
\subsection{Boundary matrix model at non-zero chemical potential}\label{canonicalboundary}
\begin{figure}[h!]
	% \begin{wrapfigure}{r}{0.43\textwidth}
	%\begin{center}
	{\centering
		\subfloat[]{\includegraphics[width=2in]{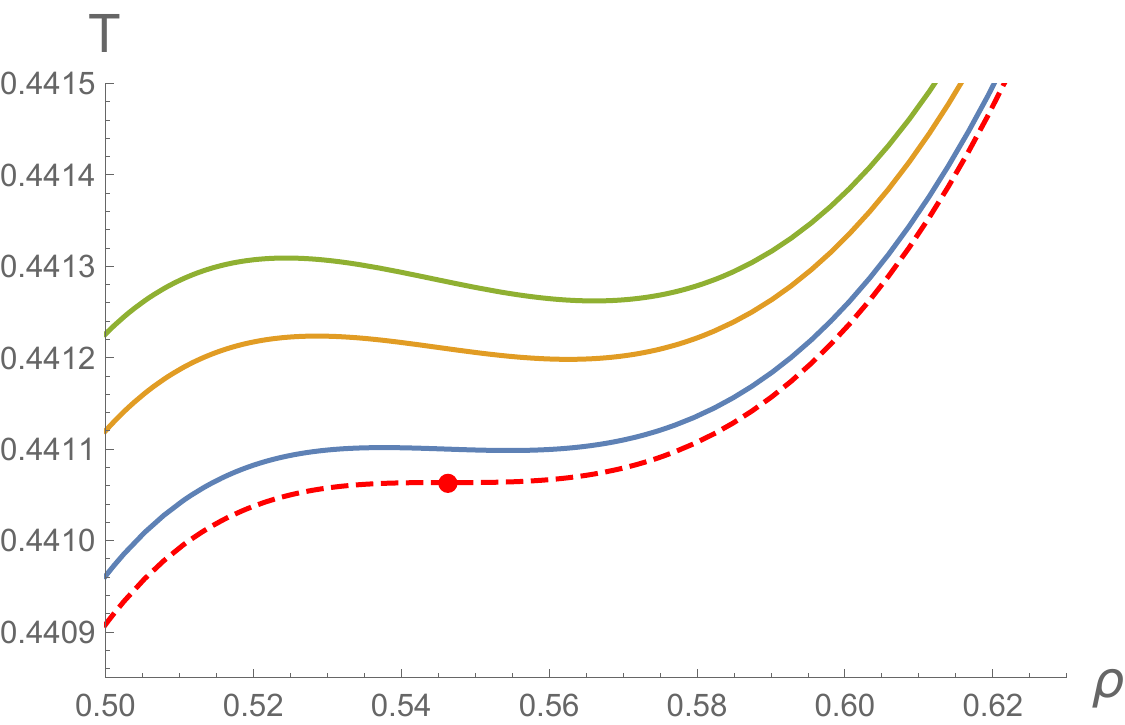}\label{fig:RN_canon_boundary_eos}}\hspace{0.2cm}
		\subfloat[]{\includegraphics[width=2in]{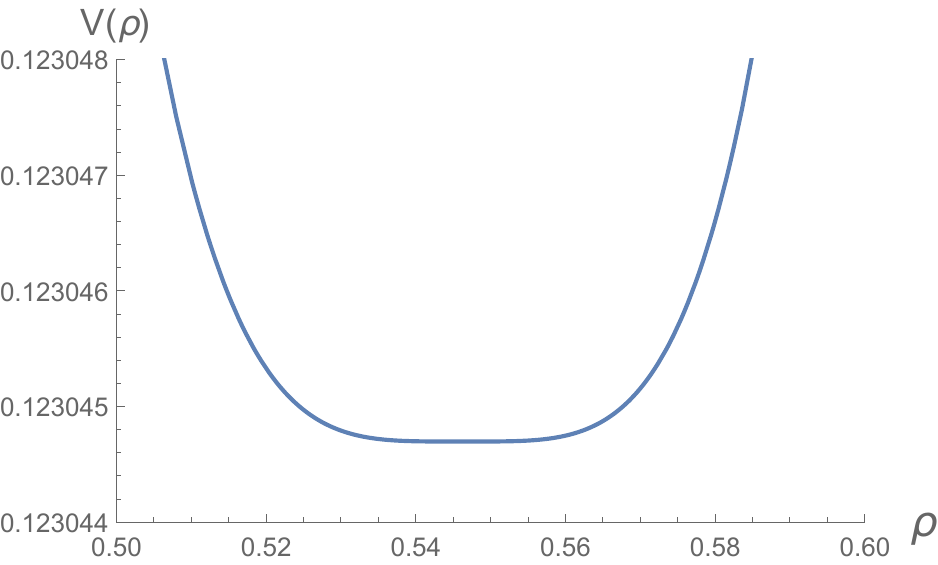}\label{Fig:RN_canon_boundary_v1_plot}}\hspace{0.2cm}	
		\subfloat[]{\includegraphics[width=2in]{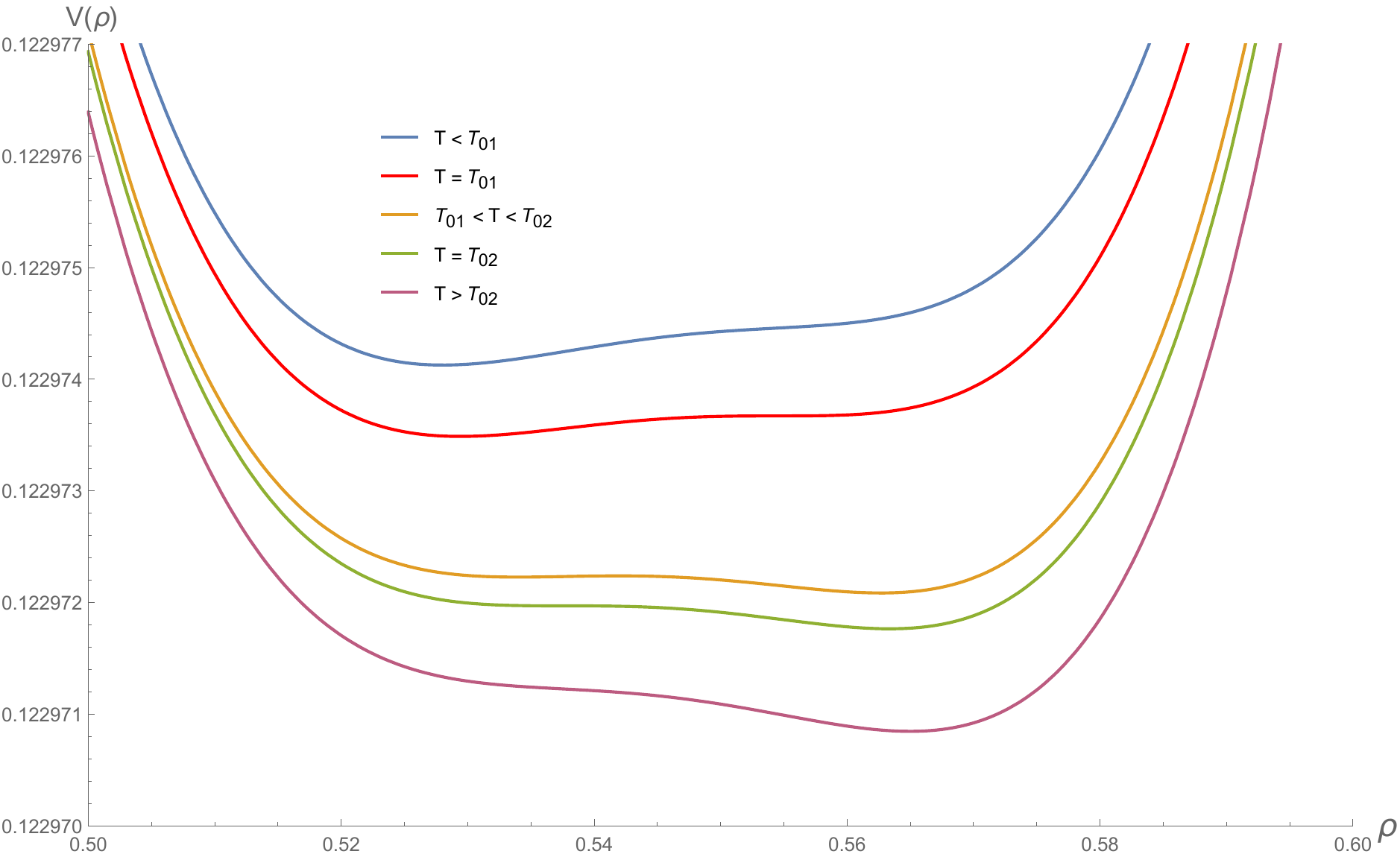}\label{Fig:RN_canon_boundary_v2_plot}}				
		% 2a,2b	
		
		\caption{\footnotesize For the $(a,b)$ matrix model with finite chemical potential (in canonical ensemble):  (a) Plot of equation of state $T(\rho, q)$ for $q \leq q_{\rm cr}$. From top to bottom, the charge increases. Dashed curve stands for the case $q=q_{\rm cr}$ with the red dot denoting the critical point; Plots in (b) and (c) are for the  potential $V(\rho)$ for the cases $ q= q_{\rm cr}$ (and  $T=T_{\rm cr}$) and for  $ q < q_{\rm cr}$, respectively.} 
	\label{Fig:RN_canon_boundary}		}
\end{figure}

%%%%%%%%%%%%%%%%%%%
One can extend the $(a, b)$ matrix model shown in eqn.~(\ref{eq: a_b_matrixmodel_for_sch}), with an additional logarithmic term in the effective potential at a
fixed nonzero charge $q$. This will serve as a boundary dual for the Reissner-Nordstrom $\text{AdS}_5$  black holes in the canonical ensemble. The  effective action of this model is~\cite{Basu:2005pj}: 
\begin{equation}
	S_q = S\big(a(T), \, b(T), \, \rho \big) + q \, \text{log}(\rho),
\end{equation} 
which is supplemented  by the following saddle point equations:
\begin{eqnarray}
	\rho F +q &=& \rho^2   \hspace{2cm} \text{for} \quad 0 \leq \rho \leq \frac{1}{2}    \\
	&=& \frac{\rho}{4(1-\rho)}  \hspace{0.8cm}  \text{for} \quad \frac{1}{2} \leq \rho \leq 1\, , \label{eq:saddle pt eq for rho >1/2 for RN boundary}
\end{eqnarray} 
where $ F(\rho) = a\rho +2b\rho^3$. The effective potential for this model is
\begin{eqnarray}
	V(\rho) &=& \frac{1-a}{2}\rho^2 -\frac{b}{2} \rho^4  -q\, \text{log}(\rho)     \hspace{4cm} \text{for} \quad 0 \leq \rho \leq \frac{1}{2}    \\
	&=& -\frac{a}{2} \rho^2 -\frac{b}{2} \rho^4 -q\, \text{log}(\rho)  -\frac{1}{4} \text{log}[2(1-\rho)] +\frac{1}{8}  \hspace{0.8cm}  \text{for} \quad \frac{1}{2} \leq \rho \leq 1. \label{eq:eff_potential for rho > 1/2-RN}
\end{eqnarray}
For the $\rho > 1/2$ region, the behaviour of the parameters $a(T)$ and $b(T)$ can be computed for $T  \geq T_{\rm cr}$ regime~\cite{Basu:2005pj,Yerra:2023hui}. The results are shown in the Fig.~\ref{fig:RN_canon_boundary_a_b_plots}.  This of course requires the usage of bulk equation of state (see the Appendix-B for details).
\begin{figure}[h!]
	
	% \begin{wrapfigure}{r}{0.43\textwidth}
	%\begin{center}
	{\centering
		
		\subfloat{\includegraphics[width=2.7in]{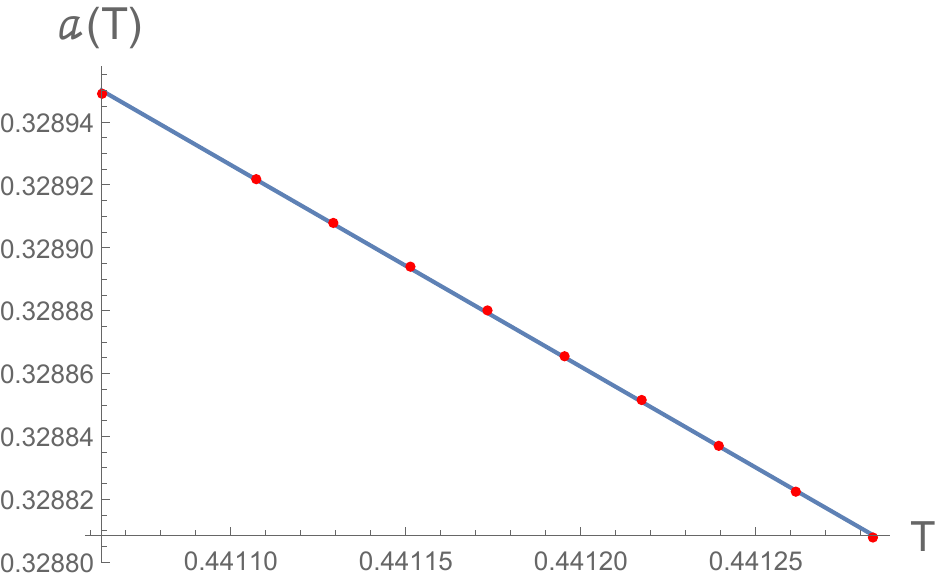}}\hspace{1.5cm}	
		\subfloat{\includegraphics[width=2.7in]{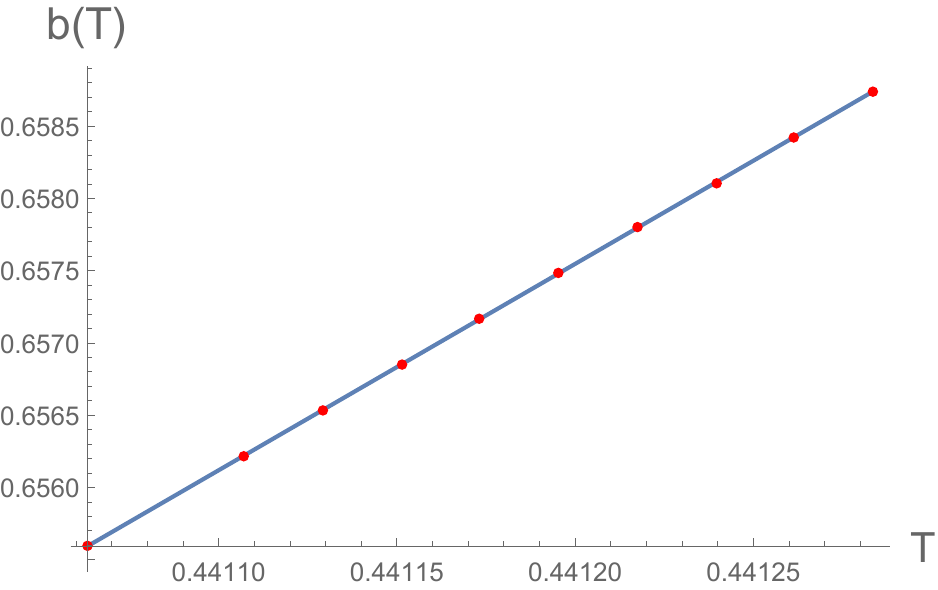}}	
		
		\caption{\footnotesize Case of $(a,b)$ phenomenological matrix model at non-zero chemical potential (in the canonical ensemble): Plots show the temperature  dependence of the parameters $a(T)$ and $b(T)$ for $ T \geq T_{\rm cr}=\frac{4\sqrt{3}}{5\pi}$. Red dots denote the data points, whose fits are represented by blue colored curves. The fitting curves are, $a(T) =c_1 T +c_2$, and$b(T) =c_3 T +c_4$. $(\text{Note:}\, c_1=-0.641573, c_2=0.611924, c_3=14.2782, c_4=-5.64198)$.}
		\label{fig:RN_canon_boundary_a_b_plots}	}
	
\end{figure}
\vskip 0.2cm
\noindent
The fitting curves for $a(T)$ and $b(T)$ can now be used to solve the saddle point equations~\ref{eq:saddle pt eq for rho >1/2 for RN boundary} for temperature, giving:
\begin{equation}
T(\rho, q) = \frac{4 q (1-\rho) + 8c_4(1-\rho)\rho^4 + 4c_2(1-\rho)\rho^2 - \rho \,}{4(\rho - 1)\rho^2 (c_1 + 2c_3 \rho^2)}.
\end{equation}
The  behaviour of the above expression extracted from curve fitting, matches qualitatively with the known bulk formula, as is evident from fig.~\ref{fig:RN_canon_boundary_eos}. The critical point can also be shown to lie at $(T_{\rm cr}, \, q_{\rm cr}, \, \rho_{\rm cr}) = (\frac{4\sqrt{3}}{5\pi}, \, \frac{1}{3\sqrt{15}}, \, 0.546293)$. 
\vskip 0.2cm
\noindent
Let us now summarize how the effective potential $V(\rho)$ at a fixed charge $q$ characterizes the phases for various  temperatures, particularly looking at figures Fig.~\ref{Fig:RN_canon_boundary_v1_plot}, and Fig.~\ref{Fig:RN_canon_boundary_v2_plot}. The stable/unstable saddle points of $V(\rho)$ (for $\rho > 1/2$) denote the corresponding stable/unstable black hole solutions in the bulk (cf. Fig.~\ref{Fig:fe3} in Appendix-B). In the regime $q < q_{\rm cr}$ at certain low temperatures, such as, $T < T_{01}$, one sees the presence of only one saddle point, which can be interpreted as the small black hole in the bulk.  A raise in temperature to $T_{01}$ results in the nucleation of two saddle points which in the bulk are the unstable intermediate and the stable large black hole. When the temperature is raised further, such that, $T_{01} <T < T_{02}$, there exist three saddle points (corresponding to small, intermediate and large black holes in the bulk). As can be seen in the aforementioned figures, for $T=T_{02}$, two of the saddle points merge (namely, the merging of small and intermediate black hole branches). Finally, for any temperature $T>T_{02}$, there is always a single saddle point, representing the stable large black hole of the bulk. There is also an interesting critical point at $q=q_{\rm cr}$, where all the three saddle points coalesce and with this, the phase structure exhibited by the effective potential $V(\rho)$ reproduces the known results from  the bulk for $q \leq q_{\rm cr}$~\footnote{ One can refer~\cite{Basu:2005pj}, for further details of the phase structure.}. We should add that $\rho = 0$ is not a solution in the non-zero charge case.
\vskip 0.2cm
\noindent
Since, close to the critical region (i.e., $q \leq q_{\rm cr}$), the effective potential $V(\rho)$  reproduces the nature of the off-shell bulk free energy, we utilize it to obtain other thermodynamic quantities as follows.
The on-shell free energy $F$, can be computed from the fitting curves for $a(T)$, $b(T)$. This is practically done by substituting $T(\rho, q)$ in  the effective potential $V(\rho)$ in equation~\eqref{eq:eff_potential for rho > 1/2-RN}, to obtain
\begin{equation}
F = -\frac{\rho^2}{2}(c_1 T + c_2) -\frac{\rho^4}{2}(c_3 T+ c_4) -q \text{log}(\rho) -\frac{1}{4}\text{log}[2(1-\rho)]+\frac{1}{8} \equiv E-TS,\label{eq:matrix_on-shell_F_charged}
\end{equation} 
\noindent
where similar to the situation in the bulk, we assume that the free energy obeys $dF=\mu dQ-SdT$~\footnote{Here, we have replaced $q$ with the physical charge $Q$, using eqn.~\eqref{eq:qQ}.}. Now, the computation of entropy $S$ of the CFT results in
\begin{equation}\label{eq:matrix_S_charged}
S=-\frac{\partial F}{\partial T} \Big\vert_{\rm Q} = -\frac{\partial F /\partial \rho}{\partial T/ \partial \rho}\Big\vert_{\rm Q}= \frac{1}{2}\rho^2(c_1+c_3\rho^2).
\end{equation}  
\noindent
Also, the energy $E$ of the CFT is 
\begin{eqnarray}
E &=& F+TS \nonumber \\
&=& \frac{1}{24}\Big( 3-12\rho^2(c_2+c_4\rho^2) - 6\text{log}[2(1-\rho)]-2\sqrt{3}Q \text{log}(\rho)  \Big),
\end{eqnarray}
where we assume that it satisfies the first law: $dE=TdS+\mu dQ$. The chemical potential $\mu$ corresponding to the charge is
\begin{equation}
\mu = \frac{\partial E}{\partial Q}\Big \vert_{\rm S}= -\frac{\text{log}(\rho)}{4\sqrt{3}}.
\end{equation}
%%%%%%%%%%
\subsection{Cardy-Verlinde formula} \label{3.2}
%%%%%%%%%%%%%
It is now possible to derive the Cardy-Verlinde (CV) formula following the set up in~\cite{cai2001cardyverlinde}. Taking the scaling of the boundary metric as in eqn.~\eqref{eq:sch_scaled_metric},
the energy $\bar{E}$, temperature $\bar{T}$, chemical potential $\bar{\mu}$, charge $\bar{Q}$, and the entropy $\bar{S}$ of the CFT are scaled as:
\begin{eqnarray}
\bar{E}&=&\frac{l}{R} E,\\
	\bar{T} &=& \frac{l}{R}T, \\
	\bar{\mu} &=& \frac{l}{R}\mu, \\
	\bar{Q}&=&Q,\\
	\bar{S}&=&S.
\end{eqnarray}
\noindent These satisfy the first law: $d\bar{E}=\bar{T}d\bar{S}+\bar{\mu}d\bar{Q}-\bar{P}d\bar{V}$. The pressure $\bar{P}$, conjugate to the volume  $\bar{V}= \omega_3 R^3$, is given by
\begin{eqnarray}
	\bar{P} &=&-\frac{\partial \bar{E}}{\partial \bar{V}} \Big \vert_{\rm \bar{S}, \bar{Q}} = -\frac{\partial \bar{E}/ \partial R}{\partial \bar{V}/ \partial R} \Big \vert_{\rm \bar{S}, \bar{Q}} \nonumber \\ 
&=& \frac{1}{72R^4}\Big( 3-12\rho^2(c_2+c_4\rho^2) - 6\text{log}[2(1-\rho)]-2\sqrt{3}Q \text{log}(\rho)  \Big).
\end{eqnarray}
\noindent We now define the Casimir energy $\bar{E}_c$ (as the violation of the Euler identity~\cite{Verlinde:2000wg,cai2001cardyverlinde}), and the electric potential energy $\bar{E}_Q$~\cite{cai2001cardyverlinde}, as:
%~\citealp*{cai2001cardyverlinde}
\begin{eqnarray}
\bar{E}_c &\equiv & 3(\bar{E}+\bar{P}\bar{V}-\bar{T}\bar{S}-\bar{\mu}\bar{Q}), \\
\bar{E}_Q &\equiv& \frac{1}{2}\bar{\mu}\bar{Q}.
\end{eqnarray}
\begin{figure}[h!]
	
	% \begin{wrapfigure}{r}{0.43\textwidth}
	%\begin{center}
	{\centering
		
		{\includegraphics[width=2.8in]{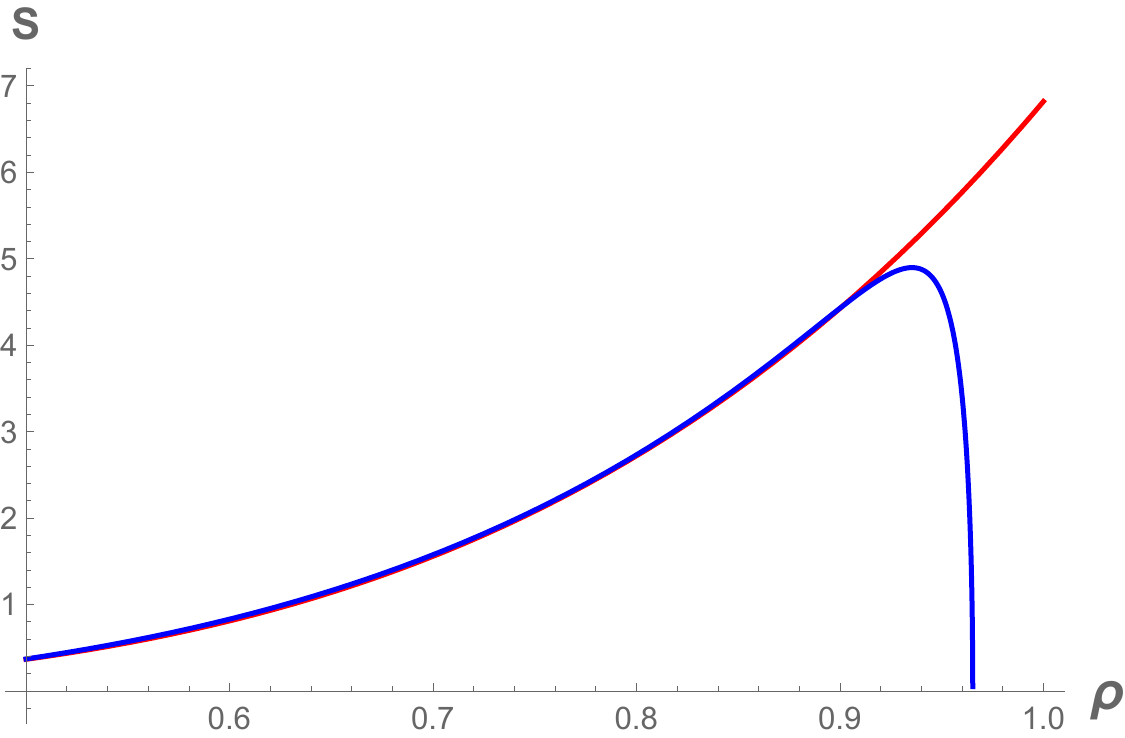}}	
		\caption{\footnotesize For $(a,b)$ matrix model with non-zero chemical potential: Red curve is for the LHS  and the blue curve is for the RHS of the equation~\eqref{eq:cv_RN}. Here, we used $Q=0.1$.}
		\label{fig:cv_RN_matrix_f}	}
	
\end{figure}
\vskip 0.2cm
\noindent
Now, as shown in the Fig.~\ref{fig:cv_RN_matrix_f}, one can numerically
%~\footnote{Here, the behaviour and the mismatch of the curves  around $\rho =1$ in the Fig.~\ref{fig:cv_RN_matrix_f}, may be due to some statistical errors and approximations considered in the model of the theory.} 
verify  the  Cardy-Verlinde formula~\cite{cai2001cardyverlinde}:
\begin{equation}\label{eq:cv_RN}
	\bar{S}=\frac{2\pi R}{3}\sqrt{\bar{E}_c [2(\bar{E}-\bar{E}_Q)-\bar{E}_c ]}.
\end{equation} 
%%%%%%%%%%%%%%%%%%%
\section{Conclusion} \label{conclusion}
%%%%%%%%%%%%%%%%%%%
To conclude, in this work, we made an attempt to verify the
Cardy-Verlinde (CV) formula directly working in the boundary gauge theory. Our computations were
facilitated by the phenomenological matrix model proposed to be
an effective model for the strongly coupled $\mathcal{N} = 4$, $SU(N)$ gauge
theory at finite temperature  and chemical potential at large $N$. Our results are summarized
in
figures~\ref{fig:cv_sch_matrix_f} and~\ref{fig:cv_RN_matrix_f}. While for small values of $\rho$, we find that the CV formula
is satisfied,
there are discrepancies when $\rho$ approaches one. Though the reasons
are not immediately obvious to us, it could be because of truncating the
model to contain terms with only a few lower powers of $\rho$ with possible statistical errors. With the inclusion of higher curvature terms in the Einstein action and in the cases involving arbitrary horizon topologies, interesting modifications of the CV formula have been noted~\cite{cai2001cardyverlinde}. Using known methods to include such corrections in the boundary matrix models~\cite{Dey:2006ds,Dey:2007vt,Dey:2008bw}, it should be a good exercise to check the validity of CV formula in more general situations.

%%%%%%%%%%%%%%%%%%%%%%%%%%%%%%%%%%%%%%%%%%%%%%%%%%%%%%%%%%%%%%%%%%%%%%%%%%%%%%%%%%
\section*{Acknowledgements}
One of us (C.B.) thanks the DST (SERB), Government of India, for financial support through the Mathematical Research Impact Centric Support (MATRICS) grant no. MTR/2020/000135. We thank IIT Bombay for warm hospitality during the Indian Strings Meeting 2023, where this work drew to a close. 

%%%%%%%%%%%%%%%%%%%%%%%
%%%%%%%%%%%%%%%%%%%%%%%%%%%%%%%%%%%%%%%% APPENDIX %%%%%%%%%%%%%%%%%%%%%%%%%%%%%%%%%%%
%%%%%%%%%%%%%%%%%%%%%%%%%%%%%%%
%\section*{Appendix A}
%%%%%%%%%%%%%\appendix
%\section{Appendices}
%\subsection{First appendix}
%\subsection{Second appendix}%%%%%%%%%%%%%%%%%%
\renewcommand{\thesection}{\Alph{section}}
%%%%%%%%%%%%%%%%%%%%%%%%%%%%%%%%%%%
%\section*{Appendix}
\appendix
\section*{Appendices}
%\addcontentsline{toc}{section}{Appendices}
%\renewcommand{\thesubsection}{\Alph{subsection}}
%\renewcommand{\thesection}{\Alph{section}\arabic{section}}
%\setcounter{section}{0}
In the following two appendices, we summarize the procedure for obtaining the off-shell free energy of Schwarzschild-AdS and the Reissner–Nordstrom-AdS black holes, and also present their phase structure, which will be helpful to refer to, when comparing with the phase structure shown by boundary matrix model in the main text.
 %%%%%%%%%%%%%%%%%%%%%%%%%%%%%%%%%%%
\section{Schwarzschild-AdS black holes} \label{AppendixA}
\begin{figure}[h!]
		
		% \begin{wrapfigure}{r}{0.43\textwidth}
		%\begin{center}
		{\centering
			
			\subfloat[]{\includegraphics[width=2.8in]{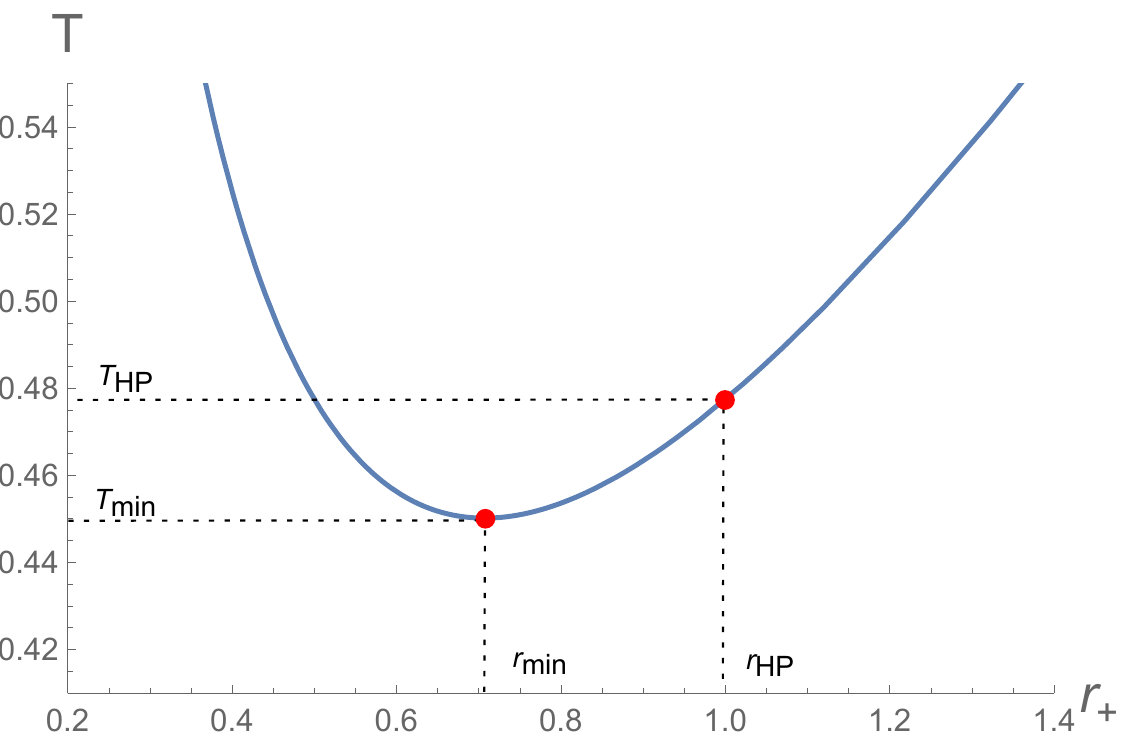}\label{fig:sch_bulk_eos}}\hspace{0.5cm}
			\subfloat[]{\includegraphics[width=2.8in]{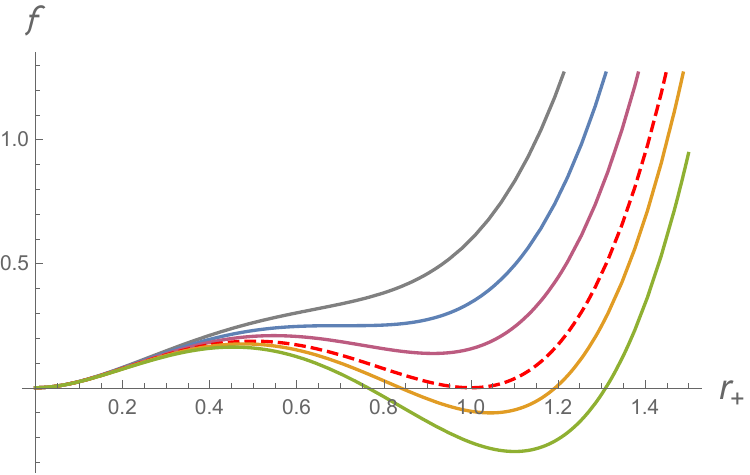}\label{fig:sch_bulk_free_energy_plot}}

			\caption{\footnotesize Schwarzschild black holes in $\text{AdS}_5$: (a) Temperature $T$  plotted in terms of the horizon radius $r_+$, showing the existence of small $(r_+ < r_{\rm min})$ and large $(r_+ > r_{\rm min})$ black hole branches for $T > T_{\rm min}$.  The Hawking-Page (HP) transition is seen at $ T = T_{\rm HP}$.  (b) This plot shows the nature of Bragg-Williams free energy $f$ with respect to the horizon radius $r_+$ at various temperatures $T$. Temperature of the curves decreases from the bottom to the top. The blue curve is at $T_{\rm min}$ and the dashed red curve corresponds to the temperature $T_{\rm HP}$. Also, $(r_{\rm min}, T_{\rm min}) = (\frac{1}{\sqrt{2}}, \frac{\sqrt{2}}{\pi})$ and  $(r_{\rm HP}, T_{\rm HP}) = (1, \frac{3}{2\pi})$.}
		\label{fig:cv_sch_bh_T}	}
		
	\end{figure}

%%%%%%%%%%%
%	\section*{Appendix}
%	\subsection*{}
Staring point is the $5$-dimensional Schwarzschild-AdS black holes in the bulk, with the action given below, followed by the line element~\cite{Witten:1998zw}:
	\begin{equation}
		I = -\frac{1}{16{\pi}G} \int d^{5}x \sqrt{-g} \left[R  +
		\frac{12}{l^2}\right],	
	\end{equation}
	\begin{equation}
		ds^2 = -V(r)dt^2 + \frac{dr^2}{V(r)} +r^{2}d{\Omega}^2_{3}, \label{eq:bulk_sch_metric}
	\end{equation}
	\noindent where $G$ denotes the Newton's constant, $d{\Omega}^2_{3}$ is the metric on the 
	unit $3$-sphere $S^3$ with volume $\omega_3$. The function $V(r)$ takes the form
	\begin{equation}
		V(r) = 1 - \frac{m}{r^2} +  \frac{r^2}{l^2}\,
	\end{equation}
	
	\noindent  with $l$ as the AdS length and $m$ related to the ADM mass $M$ of the hole.
	The usual thermodynamic quantities, such as, energy $E$, temperature $T$, entropy $S$, and finally the free energy $F$ of the hole can be written as a function of the horizon radius $r_+$ as:
	\begin{eqnarray}
		E&=&M =\frac{3\omega_{3}}{16\pi G}m  \nonumber \\
		&=&\frac{3\omega_3}{16\pi G}\big(r_+^2+\frac{r_+^4}{l^2}\big),\\
		T&=& \frac{l^2+2r_+^2}{2\pi l^2r_+}, \\
		S&=&\frac{\omega_{3}}{4G}r_+^{3},\\
		F&=& M-TS.
	\end{eqnarray}
The above quantities can be shown to satisfy the first law of thermodynamics: $dE = TdS$, with the free energy obeying $dF=-SdT$. From the equation of state i.e., the plot of temperature in Fig.~\ref{fig:sch_bulk_eos}, one deduces the existence of a minimum temperature $T_{\rm min}$, beyond which there can be two black holes, known as, the small $(r_+ < r_{\rm min})$  and large $(r_+ > r_{\rm min})$. The small black hole is locally unstable (with negative specific heat), and the larger counterpart is stable. A key point in the phase structure is at a different temperature $T_{\rm HP} > T_{\rm min}$, which corresponds to the Hawking-Page (HP) transition, beyond which the large black hole branch is preferred over othre phases. 
\vskip 0.2cm 
\noindent
The HP phase transition can be understood straightforwardly with the construction of Bragg-Williams (BW) off-shell free energy  $f$~\footnote{ Through out the manuscript, we work with the units $\hbar = c = 1$, and we set AdS length $l$, volume of the three sphere $\omega_3$, and $16\pi G$ to 1.} as~\cite{Banerjee:2010ve}:
\begin{equation} \label{fe1}
f(r_+, T)= M-TS= 3r_+^2(1+r_+^2)-4\pi r_+^3T.
\end{equation}
Here, the horizon radius $r_+$ is treated as an order parameter and the temperature $T$ as an external parameter. The nature of the free energy $f$ for various temperatures can be inferred from Fig.~\ref{fig:sch_bulk_free_energy_plot}.  
The minima of the free energy $f$, represent the locally stable large black hole phase, and the maxima represent the locally unstable small black hole phase, while the AdS space $(r_+ = 0)$ corresponds to the zero of free energy. The HP transition happens when there is a degenerate minima that satisfies the two conditions $( f=0 \, \text{and} \,  \frac{\partial f}{\partial r_+} =0)$. 
Further details of the phase structure can be found in~\cite{Hawking:1982dh,Banerjee:2010ve}.

	 %%%%%%%%%%%%%%%%%%%%%%%%%%%%%%%%%%%
\section{Reissner–Nordstrom-AdS black holes} \label{AppendixB}
\begin{figure}[h!]
			
			% \begin{wrapfigure}{r}{0.43\textwidth}
			%\begin{center}
			{\centering
				
				\subfloat[]{\includegraphics[width=2in]{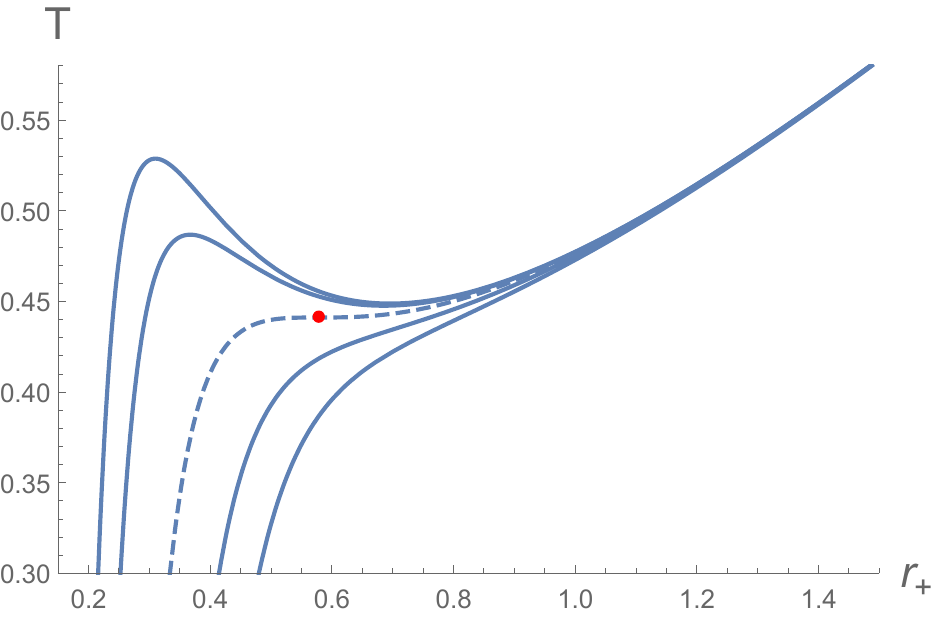}\label{fig:RN_canon_bulk_eos}}\hspace{0.2cm}
				\subfloat[]{\includegraphics[width=2in]{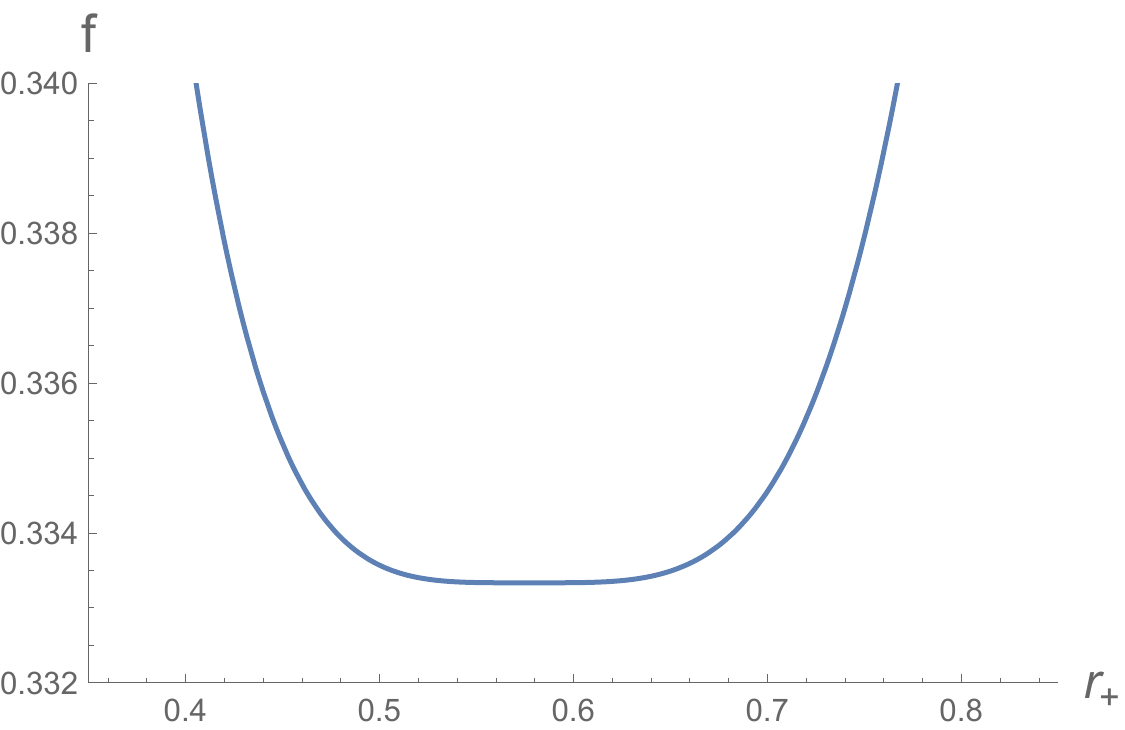}\label{fig:RN_canon_bulk_f1_plot}}\hspace{0.2cm}	
				\subfloat[]{\includegraphics[width=2in]{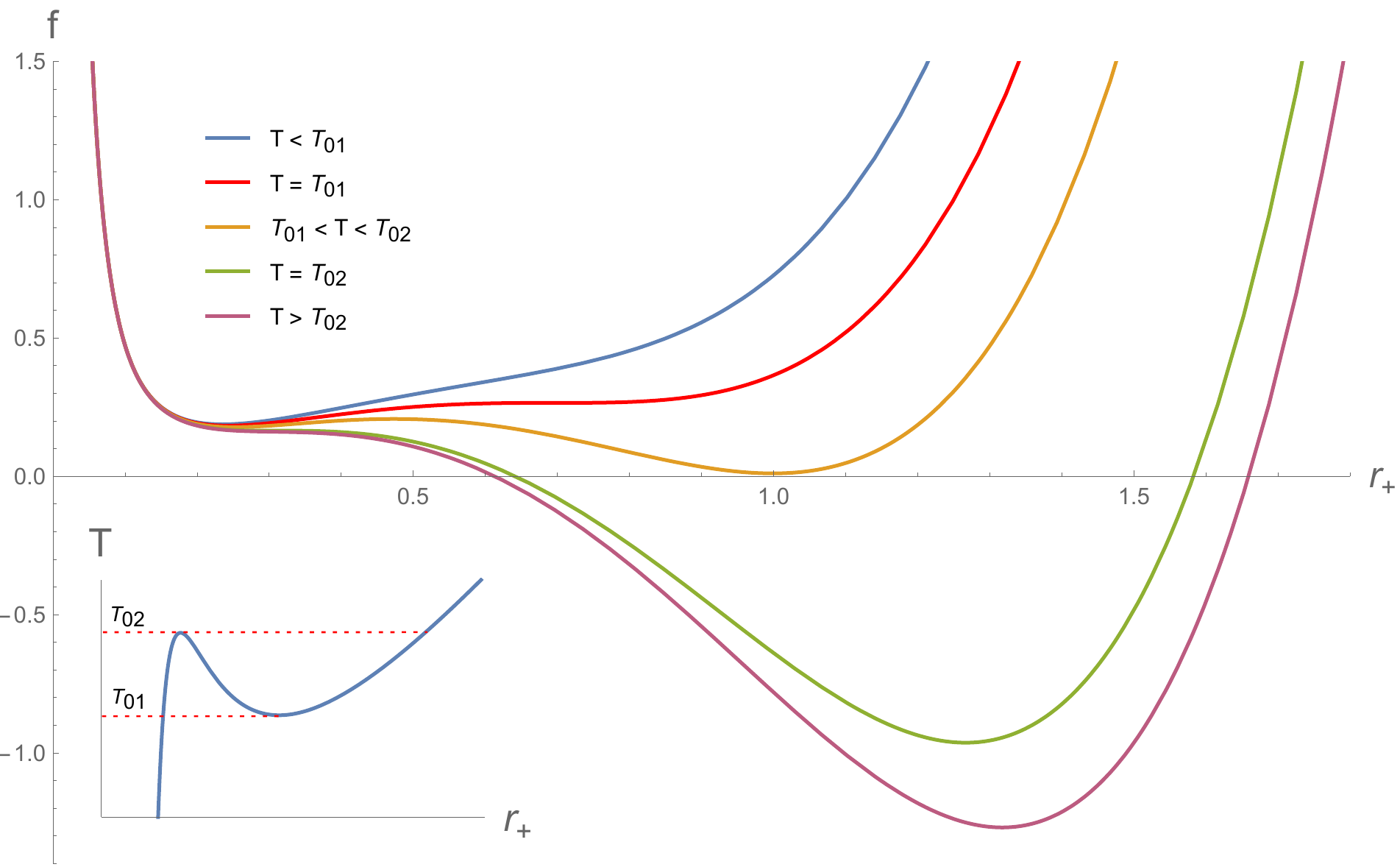}\label{fig:RN_canon_bulk_f2_plot}}
				\caption{\footnotesize The Reissner-Nordstrom black hole phases in $\text{AdS}_5$ (in the canonical ensemble): (a) Plots of equation of state, i.e., $T(r_+,q)$ for various charges $q$, inferring the presence of three black hole branches (i.e., the small, the intermediate, and the large) for $ q <q_{\rm cr}$, with one branch for $ q > q_{\rm cr}$. Charge of the curves decreases from bottom to the top. Dashed curve is plotted for $q=q_{\rm cr}$. Red dot corresponds to the critical point;  Off-shell free energy $f$ is shown for two cases, namely, (b)  $q=q_{\rm cr}$,  and  $T=T_{\rm cr}$. (c ) for $q < q_{\rm cr}$. The inset plot shows the corresponding temperatures.}
	\label{Fig:fe3}		}	
		\end{figure}

%%%%%%%%%%%
%	\subsection*{}
	Let us now consider charged black holes, namely, the  Reissner–Nordstrom-AdS black holes in $5$-dimensions in the canonical (i.e., fixed charge) ensemble. The action and the line element are known to be~\cite{Chamblin:1999tk,Chamblin:1999hg}:
	\begin{equation}
		I = -\frac{1}{16{\pi}G} \int d^{5}x \sqrt{-g} \left[R - F^2 +
		\frac{12}{l^2}\right],	
	\end{equation}
	\begin{equation}
		ds^2 = -V(r)dt^2 + \frac{dr^2}{V(r)} +r^{2}d{\Omega}^2_{3},
	\end{equation} 
	\noindent with $V(r)$ and gauge potential noted to be:
	\begin{eqnarray}
		V(r) &=& 1 - \frac{m}{r^{2}} + \frac{q^2}{r^{4}}+  \frac{r^2}{l^2}, \\
		A &= & \left(-{1\over c}{q\over r^{2}}+\mu\right)dt.
	\end{eqnarray}
	\noindent We get  $ c=2/\sqrt{3} $, and the parameter
	$q$ is related to the black hole's charge as
	\begin{equation}
		Q=2\sqrt{3}\left({\omega_{3}\over 8\pi G}\right)q. \label{eq:qQ}
	\end{equation}
	\noindent The gauge potential is picked to vanish on the horizon at $r = r_+$, which of course fixes the chemical potential
	\begin{equation}
		\mu={1\over c}{q\over r_+^{2}}.
	\end{equation}
This shows the electrostatic potential difference between the horizon and infinity.
The standard thermodynamic quantities such as, energy $E$, temperature $T$,   entropy $S$, and the free energy $F$ receive additional corrections due to the presence of charge and read as:
	\begin{eqnarray}
		E&=&M =\frac{3\omega_{3}}{16\pi G}m  \nonumber \\
		&=&\frac{3\omega_{3}}{16\pi G}\big(r_+^{2}+ \frac{q^2}{r_+^{2}} +\frac{r_+^4}{l^2}\big),\\
		T&=& \frac{2r_+^{6}+l^2 r_+^{4}-q^2l^2}{2\pi l^2r_+^{5}}, \\
		S&=&\frac{\omega_{3}}{4G}r_+^{3},\\
		F&=& M-TS.
	\end{eqnarray}
	\noindent These once again follow the first law: $ dE = TdS +\mu dQ$, and the free energy satisfies $dF=\mu dQ-SdT$.
Thermodynamics and phase transitions of charged black holes in AdS have a rich history, whose phase structure closely follows the van der Waals fluid model~\cite{Chamblin:1999tk,Chamblin:1999hg,Kubiznak:2012wp}, which is captured by the equation of state $T(r_+,q)$,  shown in  Fig.~\ref{fig:RN_canon_bulk_eos}. There is a special critical charge  $q_{\rm cr}$ and temperature  $T_{\rm cr}$, above which the black holes enter a unique stable phase. Below this critical limit, the system can exist in any of the three phases (namely,  the small, intermediate, and the large black hole branches). The small and large black hole branches are locally stable (due to positive specific heat), but the intermediate branch is unstable. There is of course a first order transition between the small black holes and the large black holes, terminating in a second order critical point, which is a point of inflection, obtained generally as
	\begin{equation}
		\frac{\partial T}{\partial r_+} = 0, \quad \text{and}, \quad \frac{\partial^2 T}{\partial r_+^2} = 0.
	\end{equation}
	This gives the critical point as $(T_{\rm cr}, \, r_{\rm cr},\, q_{\rm cr}) = (\frac{4\sqrt{3}}{5\pi}, \frac{1}{\sqrt{3}}, \frac{1}{3\sqrt{15}})$. 
		\vskip 0.2cm
		\noindent
		 Now, we can construct the Bragg-Williams off-shell free energy $f$ as in Appendix-A, where, in addition to the temperature, the charge is also an external parameter. Horizon radius $r_+$ continues to be the order parameter as before, with the free energy found to be~\cite{Chamblin:1999hg,Li:2020nsy}:
		\begin{equation}
		f = M-TS=\frac{3}{r_+^2} (r_+^6+r_+^4+q^2) - 4\pi r_+^3T.
		\end{equation}
		The phase structure of these charged black holes can now be obtained easily by studying the free energy, whose extremal points precisely correspond to the black hole solutions themselves. For instance, the minima (maxima) stand for stable (unstable) phases. For $q > q_{\rm cr}$ and at any temperature $T$, existence of a single extremal point indicates the presence of the stable large black hole phase. In addition, for $q=q_{\rm cr}$ and $T=T_{\rm cr}$, the  three extremal points merge and signal the critical point, as shown in Fig.~\ref{fig:RN_canon_bulk_f1_plot}. Further, for $q < q_{\rm cr}$ and  in the temperature range $T_{01} < T < T_{02}$, free energy $f$ contains three extremal points representing the  small, intermediate, and large black holes branches (as also emphasized in the Fig.~\ref{fig:RN_canon_bulk_f2_plot}). 
One can refer to~\cite{Chamblin:1999tk,Chamblin:1999hg,Li:2020nsy}, for further details of the phase structure, which our construction reproduces.  
%\newpage
%%%%%%%%%%%%%%%%%%%%%%%%%%%%%%%%%%%%%%%%%%%
%%%%%%%%%%%%%%%%%%%%%%%%%%%%%%%%%%%%%		
\bibliographystyle{apsrev4-1}
\bibliography{cv_matrix}
\end{document}